\documentclass[11pt, a4paper]{article}
\pdfoutput=1

\usepackage{amsmath}
\usepackage{amsfonts}
\usepackage{amssymb}
\usepackage{array}
\usepackage{latexsym}
\usepackage{graphicx}
\usepackage[english]{babel}
\usepackage{listings}
\usepackage{url}
\usepackage{bbold}
\usepackage{bm}
\usepackage{mathrsfs}
\usepackage{comment}
\usepackage{longtable}
\usepackage{cite}
\usepackage{hyperref}
\usepackage{datetime}
\usepackage[font=footnotesize,labelfont=bf]{caption} 

\usepackage{subfig}
\usepackage{graphicx}
\usepackage[countmax]{subfloat}
\usepackage[usenames,dvipsnames]{color}

\excludecomment{toexclude}
\includecomment{toinclude}

\numberwithin{equation}{section}

\hypersetup{colorlinks, citecolor=bluscuro, linkcolor=bluscuro, urlcolor=bluscuro}
\definecolor{rossos}{rgb}{0.8,0.2,0.3}
\definecolor{bluscuro}{rgb}{0.15, 0.2, .85}
\definecolor{bluchiaro}{cmyk}{1,.3,0.,0.1} 

\makeatother   
\newcommand{\GeV}{{\rm \,GeV}}
\newcommand{\TeV}{{\rm TeV}}

\newcommand{\cm}{{\rm \,cm}}
\newcommand{\s}{{\rm \,s}}

\def\de{\textrm{d}}

 \def\be   {\begin{equation}}   \def\ee   {\end{equation}}
 \def\ba   {\begin{array}}      \def\ea   {\end{array}}
 \def\bea  {\begin{eqnarray}}   \def\eea  {\end{eqnarray}}
 \def\bean {\begin{eqnarray*}}  \def\eean {\end{eqnarray*}}
 \def\nn{\nonumber}

\newcommand{\AddV}[1]{{\textcolor{black}{#1}}}

\makeatother

\begin{document}

\begin{flushright} 
SISSA  37/2014/FISI
\end{flushright}

\vspace{0.5cm}
\begin{center}

{\LARGE \textbf {
Can AMS-02 discriminate \\
[0.2cm]
the origin of an anti-proton signal?
}}
\\ [1.5cm]

{\large
\textsc{Valeria Pettorino}$^{\rm  a, d}$\footnote{\texttt{valeria.pettorino@thphys.uni-heidelberg.de}},
\textsc{Giorgio Busoni}$^{\rm b,c, }$\footnote{\texttt{giorgio.busoni@sissa.it}},
\textsc{Andrea De Simone}$^{\rm b,c, }$\footnote{\texttt{andrea.desimone@sissa.it}},
\\[0.2cm]
\textsc{Enrico Morgante}$^{\rm d, }$\footnote{\texttt{enrico.morgante@unige.ch}},
}
{\large
\textsc{Antonio Riotto}$^{\rm d, }$\footnote{\texttt{antonio.riotto@unige.ch}},
\textsc{Wei Xue}$^{\rm c,b, }$\footnote{\texttt{wei.xue@sissa.it}}
}
\\[1cm]
\large{
$^{\rm a}$
\textit{HGSFP and Institut f\"{u}r Theoretische Physik, Ruprecht-Karls-Universit\"{a}t
Heidelberg, Philosophenweg 16, 69120 Heidelberg, Germany}\\
\vspace{1.5mm}
$^{\rm b}$
\textit{SISSA, via Bonomea 265, I-34136 Trieste, Italy}\\
\vspace{1.5mm}
\
$^{\rm c}$
\textit{INFN, Sezione di Trieste, via Bonomea 265, I-34136 Trieste, Italy}\\
\vspace{1.5mm}
\
$^{\rm d}$
\textit{D\'epartement de Physique Th\'eorique and Centre for Astroparticle Physics (CAP),\\
24 quai E. Ansermet, CH-1211 Geneva, Switzerland}\\
}
\end{center}\vspace{0.5cm}

\label{firstpage}

\begin{center}
\textbf{Abstract}
\begin{quote}
Indirect searches can be used to test dark matter  models against 
expected signals in various channels, in particular antiprotons. With antiproton data available soon at  higher and higher energies, it is important to test the dark matter hypothesis against alternative astrophysical sources, {\it e.g.~}secondaries accelerated in supernova remnants. 
We investigate
the two signals from different dark matter models and different supernova
remnant parameters, as forecasted for the AMS-02, and show that they present 
a significant degeneracy.\end{quote}
\end{center}

\def\thefootnote{\arabic{footnote}}
\setcounter{footnote}{0}
\pagestyle{empty}

\newpage
\pagestyle{plain}
\setcounter{page}{1}

\section{Introduction}
\label{sec:intro}

In the years 2006-2009, the PAMELA collaboration satellite measured the flux of cosmic ray antiparticles observed on Earth. Its results \cite{PamelaPositronsExcess,PamelaAntiprotons}, also confirmed by AMS-02 \cite{AMS02positrons}, have shown a rise in the positron fraction at energies above 10 GeV . Such a rise is not  compatible with the  predictions of the standard model of cosmic rays acceleration and propagation, in which energetic protons (primaries) accelerated by astrophysical sources as SuperNova Remnants (SNR) \cite{Bell:1978zc,Blandford:1987pw} interact with hydrogen and helium nuclei of the interstellar gas, generating antiparticles (secondaries). Futhermore, 
one could  argue that an increase with energy of the positron fraction in cosmic rays most likely requires a primary source of electron-positron pairs \cite{Serpico:2008te}.

An exciting possibility is that the rise is due to Dark Matter (DM)  particles annihilating or decaying in the galactic disk, producing a flux of antiparticles that eventually reaches Earth in addition to standard cosmic rays. Such interpretation gives the interesting possibility to explain at the same time also the gamma-ray excess from the galactic center, as in Ref.~\cite{Berlin:2014pya}.
This interpretation has however some drawbacks.
First, the fact that no anomalous signal is seen in antiprotons data in the same range of energies puts severe constraints on DM properties \cite{cirellireview} and tends to favour the so-called leptophylic models, in which DM only couples to leptons. In this scenario, antiprotons data can also be used to constrain DM properties \cite{DeSimone:2013fia, fornengo_etal_2013}, since the positrons and antiprotons fluxes are correlated thanks to the electroweak corrections \cite{Ciafaloni:2010ti,Ciafaloni:2011sa,Ciafaloni:2011gv,Ciafaloni:2012gs}.
Secondly, to fit the PAMELA and the AMS-02 data with a DM model,  one usually needs a high  cross section $\left<\sigma v\right>\sim 10^{-22}\cm^3\s^{-1}$,  much higher than the reference value of $3\times10^{-26}\cm^3\s^{-1}$ expected for a stable thermal relic. In order to justify this discrepancy, one can rely on several possible, albeit ad-hoc,  explanations: introduce a boost factor, possibly due to clumpiness of the dark matter halo \cite{Lavalle:2006vb,Lavalle:1900wn} or to the presence of a narrow resonance just below the threshold \cite{Cirelli:2008pk,Ibe:2008ye,Feldman:2008xs}; invoke non-perturbative effects operating at small velocities that can enhance the present day thermal cross section \cite{Sommerfeld,Hisano:2003ec,Hisano:2004ds,Hisano:2006nn,Cirelli:2007xd,Cirelli:2008id,ArkaniHamed:2008qn,Lattanzi:2008qa} or otherwise discard the standard thermal relic picture for DM particles.

Of course, one may invoke astrophysical sources as an explanation for the positron rise. It has been  known since a long time that a rise in the positron fraction can be due to the production of $e^\pm$ in pulsars \cite{Aharonian:1995zz}. In particular, young nearby pulsars plus a diffuse background of mature pulsars can fit PAMELA positrons data \cite{Hooper:2008kg,Grasso:2009ma}. The intrinsic degeneracy between the pulsar and the  DM interpretation of PAMELA and AMS-02 data cannot be broken by positron data alone \cite{Pato:2010im}; nevertheless the two scenarios can be distinguished by a future positive signal in the antiprotons channel since antiprotons are not expected to arise from pulsars.

Given the forthcoming release of the antiproton data from the AMS-02 collaboration, it is legitimate
to ask whether a possible antiproton signal above the expected background would lead to a degeneracy problem between   a possible DM origin and an astrophysical origin. As a benchmark model for the astrophysical source of antiprotons we take the one  discussed in  Ref. \cite{Blasi:2009hv} to explain the rise of positrons and subsequently in Ref. \cite{Blasi:2009bd} to predict the antiproton flux. The excess of positrons is due to secondary products of hadronic interactions inside the same SuperNova Remnants (SNR) that accelerate cosmic rays. Primary protons accelerated in shock regions of SNRs can undergo hadronic interactions not only at late times after diffusion in the galaxy, but also when they are still in the acceleration region. These interactions will produce a flux of antiparticles that will in turn be accelerated by the same sources of the standard primary cosmic rays, and will then give an additional cosmic ray flux at Earth with a spectral shape different from that of standard secondaries. A generic prediction of the model is a flattening and eventually a weak rise of the antiparticle-over-particle ratio in both positrons and antiprotons channel \cite{Blasi:2009bd}.
What makes this mechanism particularly interesting is that it does not need any new source of antiparticles  (since positrons and antiprotons are generated by the same primary protons that accelerate in SNR) and that it predicts similar signals both in positrons and in antiprotons, precisely as many DM model  do. This  leads to a possible degeneracy in the shape of signals of very different origin, thus weakening the  discriminating power of AMS-02.

The goal of this paper is  precisely to study this possible  degeneracy by using  the projected sensitivity of AMS-02 for the antiproton  channel under the assumption
that the measurements of AMS-02 will  show a significant antiproton excess above the background. We will assume in turn that this excess is due either to DM annihilation or to 
SNR and investigate whether the signal can be mimicked by SNR and DM annihilation, respectively. Our conclusions will be pessimistic: the expected sensitivity of an experiment like
AMS-02 may not be able to disentangle the two possible sources.

The paper is organized as follows. 
In Section \ref{sec:SNR} we review the basics of the mechanism for primary antiprotons from SNR
and recall some results which will be used in the following. 
In Section \ref{sec:secondary}, some standard material about the background of secondary antiprotons
and their propagation is recalled, while in Section \ref{sec:DM} we briefly discuss the possible
antiproton contribution from DM.
Then, in Section  \ref{sec:DMtoSNR} we turn to investigate the degenercies which may arise in the interpretation of a putative
signal in antiprotons eventually measured by AMS-02. We first assume the signal is due to DM
and we try to fit it with SNR, and subsequently we analyse briefly the possibilty of a SNR signal intepreted as a DM. Finally, our conclusions are summarized in Section  \ref{sec:conclusions}.

\section{Antiprotons accelerated in supernova remnants}
\label{sec:SNR}

 Here we briefly recall the basics of the astrophysical mechanism leading to primary antiprotons and we  refer to the original papers, Refs.~\cite{Blasi:2009hv,Blasi:2009bd}, for further details. In particular,
 Ref.~\cite{Blasi:2009bd} derived the analytical prescription for the ratio $\bar{p}/p$ that we will use for our analysis. Simulations were also performed in Ref. \cite{Kachelriess:2011qv}.

Antiproton production inside the accelerator is described by the source function
\be
Q_{\bar p}(E) = 2 \int_E^{E_{\rm max}} \de \mathcal{E} N_{\rm CR}(\mathcal{E}) \sigma_{p\bar p}(\mathcal{E},E) n_{\rm gas} c,
\ee
where $c$ is the speed of light, $N_{\rm CR}$ is the spectrum of protons inside the source, $n_{\rm gas}$ is the gas density in the shock region and $\sigma_{p\bar p}(\mathcal{E},E)$ is the differential cross section for a proton of energy $\mathcal{E}$ to produce an antiproton of energy $E$ in $pp$ scattering, that we parametrize as in Refs.~\cite{Tan:1982nc,Tan:1983de, Bringmann:2006im}. 

The energy $E_{\rm max}$ is the maximum energy of a proton accelerated in the SNR at the age relevant for this mechanism. \AddV{We will treat $E_{\rm max}$ as a free parameter in our analysis.}
The factor of 2 comes from the fact that, in $pp$ collisions, an antineutron can be produced with equal probability than an antiproton (in the isospin symmetry limit); \AddV{they will} then decay into an antiproton, contributing equally to the final flux. \AddV{For that, we} are assuming that the characteristic size of the SNR is larger than the mean path travelled by a neutron before decay.

After being produced, the antiprotons undergo acceleration around the shock region. The $\bar p/p$ flux ratio at this stage is \cite{Blasi:2009bd}
\be\label{eq:SNRratio}
\left.\frac{J_{\bar p}(E)}{J_p(E)}\right|_{\rm SNR} \sim 2\, n_1\, \epsilon \,c \left[\mathcal{A}(E)+\mathcal{B}(E)\right],
\ee
where
\be\label{eq:Aterm}
\mathcal{A}(E) = \gamma \left(\frac{1}{\xi}+r^2\right) \int_m^E \de\omega\, \omega^{\gamma-3} \frac{D_1(\omega)}{u_1^2}
\int_\omega^{E_{\rm max}} \de\mathcal{E} \,\mathcal{E}^{2-\gamma} \sigma_{p\bar p}(\mathcal{E},\omega)
\ee
and
\be\label{eq:Bterm}
\mathcal{B}(E) = \frac{\tau_{\rm SN} r}{2 E^{2-\gamma}}
\int_E^{E_{\rm max}} \de\mathcal{E}\, \mathcal{E}^{2-\gamma} \sigma_{p\bar p}(\mathcal{E},E).
\ee
The two terms $\mathcal{A}$ and $\mathcal{B}$ account for the antiparticles that are produced in the acceleration region and for the ones that are produced in the inner region of the SNR.
In the above expressions, $n_1$ and $u_1$ are the background gas target density and the fluid velocity in the upstream region of the shock, fixed as in Ref. \cite{Blasi:2009bd}  to $2\cm^{-3}$ and $0.5\times10^{-8}\cm/{\rm s}$, respectively.

The factor $\xi$ in the $\mathcal{A}$ term gives the fraction of proton energy carried away by the produced secondary antiproton, which is here taken to be constant with energy. The validity of this assumption is discussed in Ref. \cite{Kachelriess:2011qv}. In this work, we keep it as a constant and we consider it as a second free parameter for our analysis.

Both $\mathcal{A}$ and $\mathcal{B}$ include $r$, which is the compression factor of the shock, defined as the ratio of the fluid velocity upstream and downstream, and $\tau_{\rm SN}$ is the typical SNR age. 
The index $\gamma$ gives the slope of the spectrum in momentum space, and it is related to the shock compression factor by $\gamma=3r/(r-1)$. As we aim at comparing the SNR $\bar{p}/p$ ratio with the ones generated by DM annihilation, our choice is to make sure that our choice for $r$ is consistent with the ones for the background antiproton spectrum (see also discussion below) and satisfies the relation $r = (2 + \gamma_{\rm pr})/(\gamma_{\rm pr}-1)$, where $\gamma_{\rm pr} = 2 - \gamma$ is the nuclei source spectral index for the Cosmic Ray (CR) propagation model, as defined in Ref.~\cite{Evoli:2011id}; we then fix $r = 3.22$, which is consistent with $\gamma_{\rm pr} = 2.35$ of both KRA and THK models of propagation (cf.~Table \ref{tab:propagation}).

The $\epsilon=1.26$ factor in front of Eq. (\ref{eq:SNRratio}) accounts for the fact that $\bar p$ production happens not only in $pp$ collisions, but also in collisions with heavier nuclei, depending on the chemical composition of the gas and it is fixed as in Ref. \cite{Blasi:2009bd}. 
The diffusion coefficient upstream the shock $D_1$ is given by
\be
D_1(E) = \left(\frac{\lambda_c c}{3 \mathcal{F}}\right) \left(\frac{E}{e B \lambda_c}\right)^{2-\beta},
\ee
where, using the same notation as in \cite{Blasi:2009bd}, $e$ is the unit charge, $B$ is the magnetic field, $\mathcal{F}\sim(\Delta B/B)^2$ is the ratio of power in turbulent magnetic field over that in the ordered one, $\lambda_c$ is the largest coherence scale of the turbulent component, and $\beta$ is the index that characterizes the spectrum of $B$ fluctuations.
Following Ref.~\cite{Blasi:2009bd} we assume a Bohm-like diffusion index $\beta=1$ and set $\mathcal{F}=1/20$ and $B=1{\,\mu{\rm G}}$. In this way the expression for $D_1$ symplifies to
\be
D_1(E) \simeq 3.3\times20\times10^{22} \,E_\GeV \cm^2 \s^{-1}.
\ee
Note that this diffusion coefficient can be different from the one assumed in propagating particles through the galaxy, since it refers only to the acceleration region near the shock.
Instead, diffusion in the galaxy affects in the same way both primary protons and antiprotons, so that the modifications in their spectra cancel out in the ratio. The flux ratio on Earth is then given by 
Eqs.~(\ref{eq:SNRratio}), (\ref{eq:Aterm}) and (\ref{eq:Bterm}).
 All in all, we have used all parameters as in Ref. \cite{Blasi:2009bd}, except $r$ (which, again, is chosen to be consistent with our choice of the propagation model). 
 
 As for the  parameters to vary in our following analysis, we have chosen  $E_{\rm max}$ and $\xi$:  we checked that they  are   the parameters  having  the largest impact on our estimate of the flux ratio. 
We have solved the equations above numerically in order to estimate the ratio (\ref{eq:SNRratio}) and we have  checked that our results match the ones in Ref. \cite{Blasi:2009bd} for the same choice of parameters. Note that for the analysis we have not  used the expression for the background illustrated in Ref. \cite{Blasi:2009bd},  but rather the one obtained from the DRAGON\cite{Evoli:2008dv} numerical code, as illustrated in the next section. Finally, we have neglected energy losses, which are not relevant for antiprotons, and solar modulation, which has negligible  effect for $E \gtrsim 10\GeV$, to which we restrict our analysis.

\section{Secondary antiprotons}
\label{sec:secondary}

As summarized above, the standard source of antiprotons in cosmic rays is the spallation of primary protons (\emph{i.e.} protons accelerated in SNR) with nuclei of the interstellar medium (ISM).
In a scenario in which the mechanism outlined in section \ref{sec:SNR} is operative, the total antiproton flux ratio would be given by the secondary component computed in this section, plus the primary component given by Eq.  (\ref{eq:SNRratio}).

In general, the propagation of Cosmic Rays through the galaxy is regulated by the diffusion equation (see for instance Ref. \cite{Evoli:2008dv})
\bea
\frac{\partial N_i}{\partial t}-{\bf\nabla}\cdot(D{\bf\nabla}-{\bf v}_c)N_i + \frac{\partial}{\partial p}\left(\dot p - \frac{p}{3}{\bf\nabla}\cdot{\bf v}_c\right) N_i
- \frac{\partial}{\partial p}p^2D_{pp} \frac{\partial}{\partial p} \frac{N_i}{p^2} = \nn \\
= Q_i(p,r,z) + \sum_{j>i} v\,n_{\rm gas}(r,z)\sigma_{ij}N_j - v\,n_{\rm gas}\sigma_i^{\rm in}(E_k)N_i,
\label{eq:diffusion}
\eea
where  $N_i(p,{\bf x})$ is the number density of the $i$-th nuclear species, $p$ is its momentum (not to be confused with the symbol for the proton) and $v$ its velocity. $D$ is the diffusion coefficient in the galaxy in real space, while $D_{pp}$ is the diffusion coefficient in momentum space, that describes the diffusive reacceleration of CRs in the turbulent galactic magnetic field. The cross sections $\sigma_i^{\rm in}$ and $\sigma_{ij}$ are the total inelastic cross section onto the ISM gas and the cross section for production of species $i$ by fragmentation of species $j$, respectively. $E_k$ is the kinetic energy of the particle under consideration. The ISM gas density is given by $n_{\rm gas}$ and  ${\bf v}_c$ is the convection velocity. Finally, $Q_i(p,r,z)$ is the source function that describes the injection of primary CRs in the galaxy.
\AddV{The diffusion coefficients are parametrized} as
\be
D(\rho,R,z) = D_0 \left(\frac{v}{c}\right)^\eta e^{|z|/z_t} \left(\frac{\rho}{\rho_0}\right)^\delta
\ee
and
\be
D_{pp} = \frac{4}{3\delta(4-\delta^2)(4-\delta)} \frac{v_A^2 p^2}{D},
\ee
where $(R,z)$ are the usual cylindrical coordinates, $z_t$ is the half-height of the cylindrical diffusion box, $\rho=pv/(Ze)$ is the particle rigidity and $v_A$ is the Alfv\'en velocity.

To compute the secondary antiproton flux, we have assumed a spectrum of primary protons from SNR of the form $Q_p\sim \rho^{-\gamma_{\rm pr}}$, and  then solved the diffusion Eq. (\ref{eq:diffusion}) numerically using the public avaiable DRAGON code \cite{Evoli:2008dv}.

In the present paper, we have considered two propagation models, namely KRA and THK, defined from the choice of propagation parameters and injection spectra illustrated in Table II of Ref. \cite{Evoli:2011id}, found by looking for good fits to B/C data and PAMELA proton data. We report the values in Table \ref{tab:propagation} for convenience. 
We have not considered other propagation models here, as we expect different choices will not 
change dramatically our main conclusions.

To constrain DM models and some SNR parameters, 
the antiproton ratio data with energy larger than $10~\GeV$ is applied.
Since the relative high energy, 
solar modulation and the
factors $\eta$ and $v_A$ in the propagation models do not play important role.

\begin{table}[htb]
\begin{center}
\begin{tabular}{|c|c|c|c|c|c|c|c|}
\hline
Model & $z_t$ & $\delta$ & $D_0 (10^{28} \cm^2\s^{-1})$ & $\eta$ & $v_A ({\rm km}\s^{-1})$ & $\gamma$ & $v_c$ \\
\hline
KRA & $4$ kpc & $0.50$ & $2.64 $ & $-0.39$ & $14.2\,$ & $2.35$ & $0$ \\
\hline
THK & $10$ kpc & $0.50$ & $4.75$ & $-0.15$ & $14.1\,$ & $2.35$ & $0$ \\
\hline
\end{tabular}
\caption{Diffusion parameter values used to propagate the secondary antiproton flux and the DM originated flux. No solar modulation is included.}
\label{tab:propagation}
\end{center}
\end{table}
%

\section{Antiprotons from DM}
\label{sec:DM}
The production of CR's by DM annihilation is controlled by three factors: the density of DM particles in the galaxy, the details of the annihilation process (annihilation channel and fragmentation functions) and finally propagation to Earth.
The DM density profile of the Milky Way is rather uncertain, and this fact reflects in an uncertainty of ${\cal O}(\lesssim1)$ order of magnitude in the resulting flux at Earth \cite{Cirelli:2013hv}.
As a reference DM halo density profile, we have used the Navarro-Frenk-White (NFW) \cite{Navarro:1995iw} profile 
\be
\rho_{\rm NFW}(r) = \frac{\rho_s}{(r/r_s) (1 + r/r_s)^2},
\ee
with  $r_s=24.42{\rm\,kpc}$ and $\rho_s= 0.184 \GeV\cm^{-3}$ and the isothermal profile \cite{iso}

\be
\rho_{\rm ISO}(r) = \frac{\rho_s}{1  + (r/r_s)^2},
\ee
with  $r_s=4.38 {\rm\,kpc}$ and $\rho_s= 1.387 \GeV\cm^{-3}$. The propagation of cosmic rays is still controlled by Eq. (\ref{eq:diffusion}), with the source term $Q_{\bar p}$ now given by
\be
Q_{\bar p}(\vec{r},t,p) = \frac{1}{2} \left(\frac{\rho_{\rm DM}({\vec r})}{m_{\rm DM}}\right)^2 \frac{\de N_{\bar p}}{\de E} \left<\sigma v\right>,
\ee
where $\left<\sigma v\right>$ is the DM annihilation cross section and $\de N_{\bar p}/\de E$ is the number of antiprotons of a given energy $E$ per DM annihilation.
We have computed the antiproton flux at Earth using DRAGON \cite{Evoli:2011id} for various models of annihilating DM, as summarized in Table \ref{tab:result} and  including electroweak corrections \cite{Ciafaloni:2010ti}. The models have been  chosen so that they are not excluded by present antiproton data \cite{Cirelli:2013hv}. The diffusion  parameters are still the ones given in Table 
\ref{tab:propagation}.

In calculating the flux we include secondary antiprotons obtained from the scattering of primary proton with the interstellar gas.

\begin{table}[htb]
\begin{center}
\begin{tabular}{|c|c| c| c | c |c | c }
   \hline
Name &      Final state & Propagation model & DM mass ($\TeV$) 
         & $ \sigma v_0$ ($ {\rm cm}^3/ {\rm s}$) & Profile     \\
   \hline
   \hline
bKN &    $ b\bar{b}$ & KRA & 3 &  $ 7 \times 10^{-25}$   &  NFW    \\
   \hline
muKN &     $\mu^+\mu^-$& KRA & 4 &  $ 8 \times 10^{-23}$   &  NFW    \\
   \hline
muKI &    $\mu^+\mu^-$& KRA & 4 &  $ 1 \times 10^{-22}$   &  ISO  \\
   \hline
 WKN &   $W^+W^-$& KRA & 3 &  $ 7 \times 10^{-25}$   &  NFW    \\
   \hline
      \hline
 bTN &    $b\bar b$ & THK & 3 &  $ 7 \times 10^{-25}$   &  NFW    \\
   \hline
  muTN  & $\mu^+\mu^-$& THK & 4 &  $ 8 \times 10^{-23}$   &  NFW    \\
   \hline
  muTI &  $\mu^+\mu^-$& THK & 4 &  $ 1 \times 10^{-22}$   &  ISO    \\
   \hline
  WTN & $W^+W^-$& THK & 3 &  $ 7 \times 10^{-25}$   &  NFW   \\
   \hline
\end{tabular}   
   \caption {DM annihilation models considered in this analysis.
}   \label{tab:result}
\end{center}
\end{table}

\section{Investigating the  degeneracies: fit DM signal using SNR model}
\label{sec:DMtoSNR}

Our aim is to test whether a putative signal in the ratio of $\bar{p}/p$ eventually observed by AMS-02 leads to degeneracies in the interpretation of its origin:  DM or astrophysics?  To this end, 
we produce a set of mock AMS-02 data through a set of benchmark DM models and ask if these data could  be interpreted as due to SNR, based on the astrophysical mechanism described in section \ref{sec:SNR} (and using the same propagation model).

As we mentioned already, we consider as  free parameters in the SNR model  the fraction 
of proton energy carried away by the antiproton $\xi$, and the energy cutoff
$E_{\rm max}$. In order to investigate possible degeneracies, we have performed the following steps:
\vskip 0.2cm
\begin{itemize}
\item obtain the CR background expected for $\bar{p}/p$ using DRAGON, as described in section \ref{sec:secondary}; 
\item produce mock data for AMS, as described in the following; 
\item create a grid in the plane ($E_{\rm max}, \xi$), in a range of values of $ 1$ TeV $ < E_{\rm max} < 10$ TeV and $0.1 < \xi < 0.5$ \cite{Blasi:2009hv, Blasi:2009bd}; 
\item  solve Eq.~(\ref{eq:SNRratio}) numerically in order to get the ratio of $\bar{p}/p$ from SNR, as described in section \ref{sec:SNR} on the grid, assuming the same cosmic ray background as the one used for DM models;  
\item  calculate the $\chi^2$, summed on each bin for a given mock dataset, between the DM mock flux and the SNR flux. We have performed this calculation on every point of the grid to get a function $\chi^2(E_{\rm max}, \xi)$; 
\item  estimate the minimum of the $\chi^2$ for each mock dataset. Then, assuming a Gaussian distribution, the confidence contours in the plane $(E_{\rm max}, \xi)$ are plotted. The area within the contours will give us a measure of the degeneracy between DM and SNR interpretation of the mock data.
\end{itemize}
To create the mock data, we have considered a series of benchmark (fiducial) DM models and calculate the corresponding mock data for all of them, assuming a propagation method for Cosmic Rays (KRA or THK) and a DM halo profile.
In particular, we have studied  non-relativistic DM annihilating into two standard model (SM) fermions or gauge bosons with $100 \%$ 
branching ratio, such as $\chi \chi \rightarrow b  \bar{b} $, $\chi \chi \rightarrow \mu^+ \mu^- $,
and $\chi  \chi \rightarrow W^+ W^-$.
Their cross sections are chosen in such a way that they are 
consistent with the current PAMELA antiproton flux \cite{PamelaAntiprotons} and
also not excluded by the other indirect detection observations: 
the positron fraction from  PAMELA \cite{PamelaPositronsExcess} and AMS-02
\cite{AMS02positrons}, Fermi LAT's gammay ray observation of dwarf galaxies
\cite{Ackermann:2013yva} and  diffuse background \cite{Ackermann:2012rg}.
The DM benchmark models with different final states, annihliation cross section
and density profiles are listed in Tab.~\ref{tab:result}.

\begin{figure*}
\begin{center}
\includegraphics[width=0.6\textwidth]{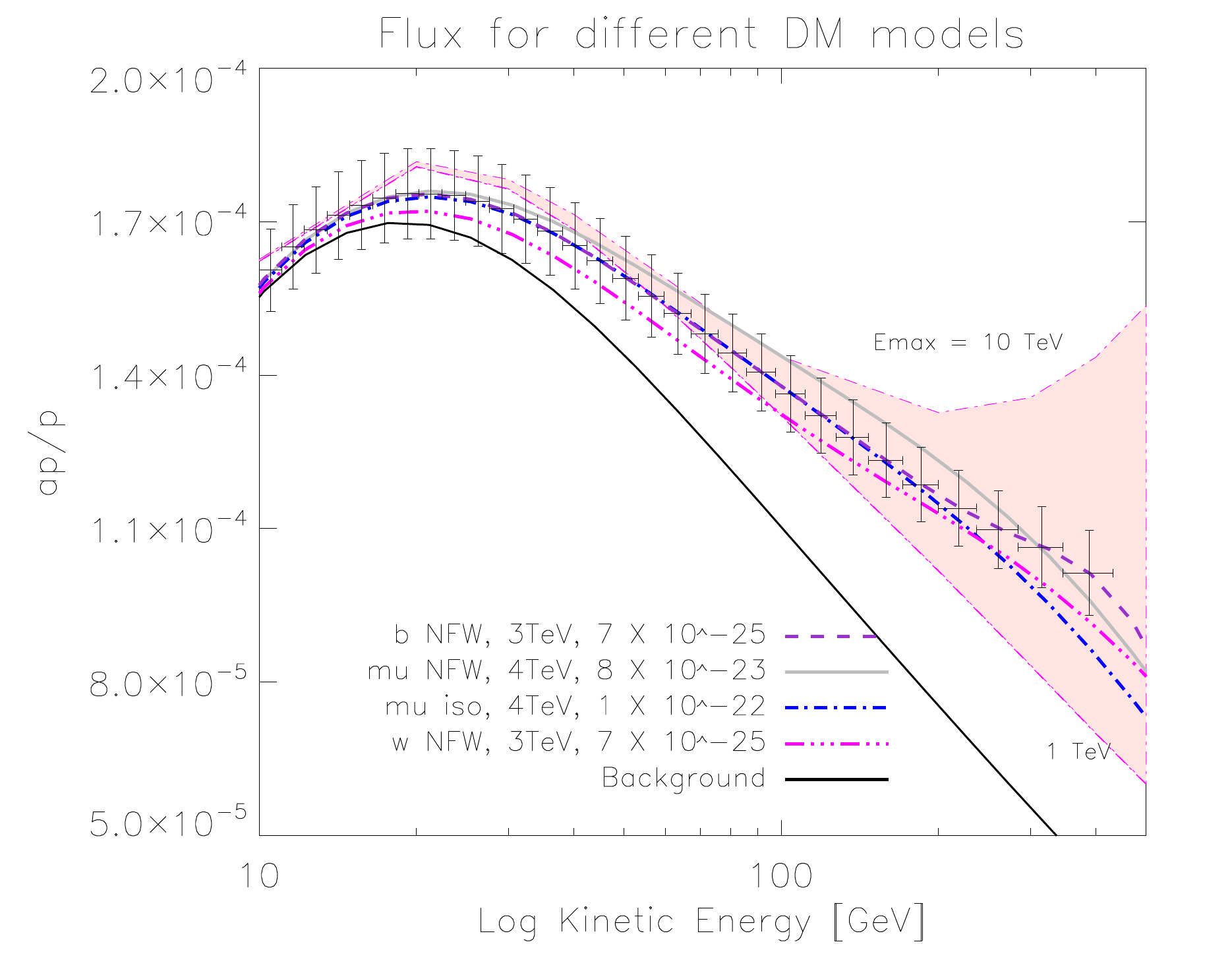}
\caption{The flux of $\bar{p}/p$ is plotted as a function of the kinetic energy for different DM models. 
The labels in the legend refer to annihilation channel, the DM halo profile, DM mass and annihilation cross section
(in units of cm$^3$/s), respectively.
The background from Cosmic Rays is shown in solid black line. For the first model we also overplot the corresponding mock data. The pink band corresponds to the region spanned by SNR when $\xi = 0.17$, as in \cite{Blasi:2009hv} and 1 TeV $< E_{\rm max} <$ 10 TeV. The propagation model used  is KRA.
}
\label{fig:mock_kra}
\end{center}
\end{figure*}

\begin{figure*}
\begin{center}
\includegraphics[width=0.6\textwidth]{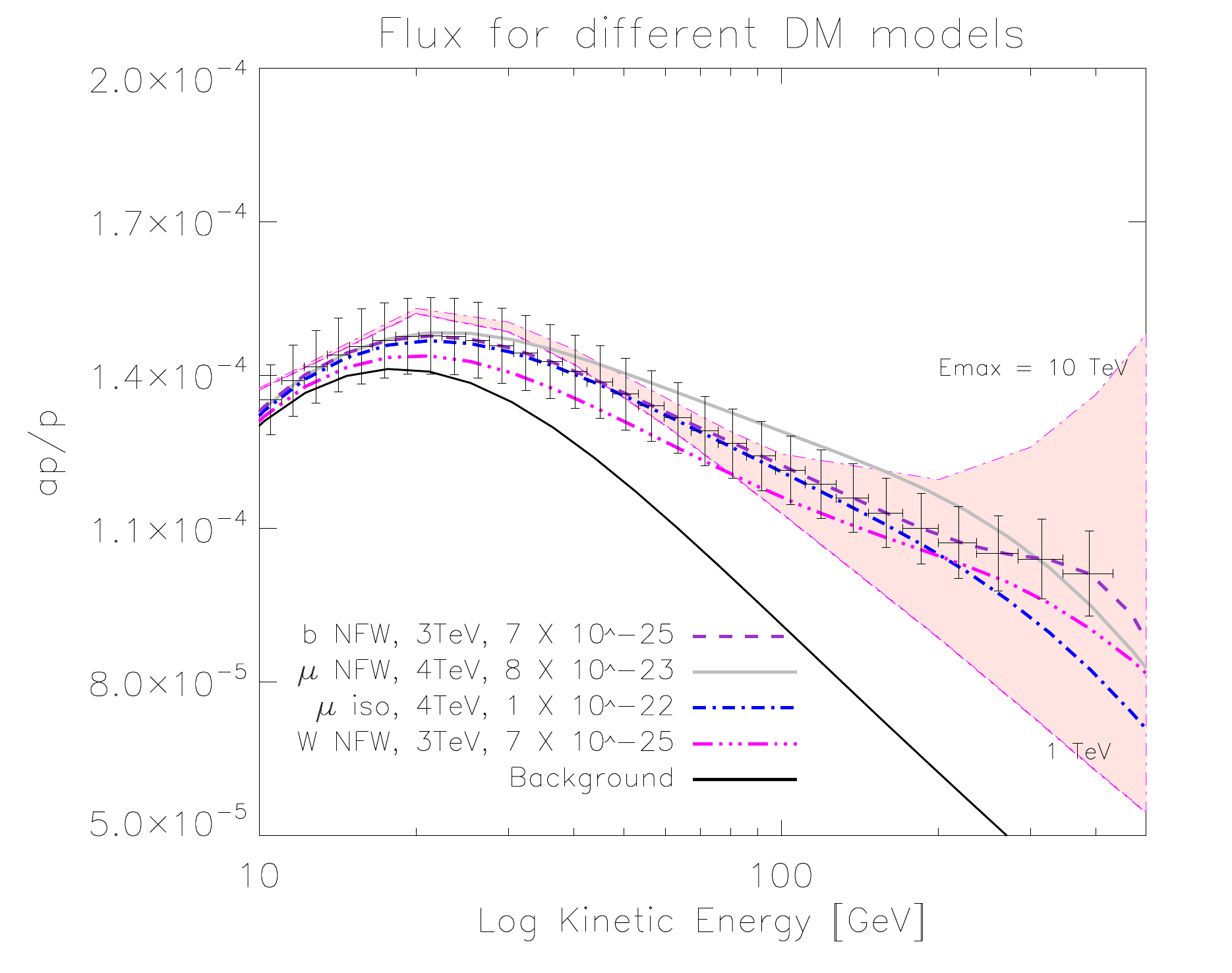}
\includegraphics[width=0.6\textwidth]{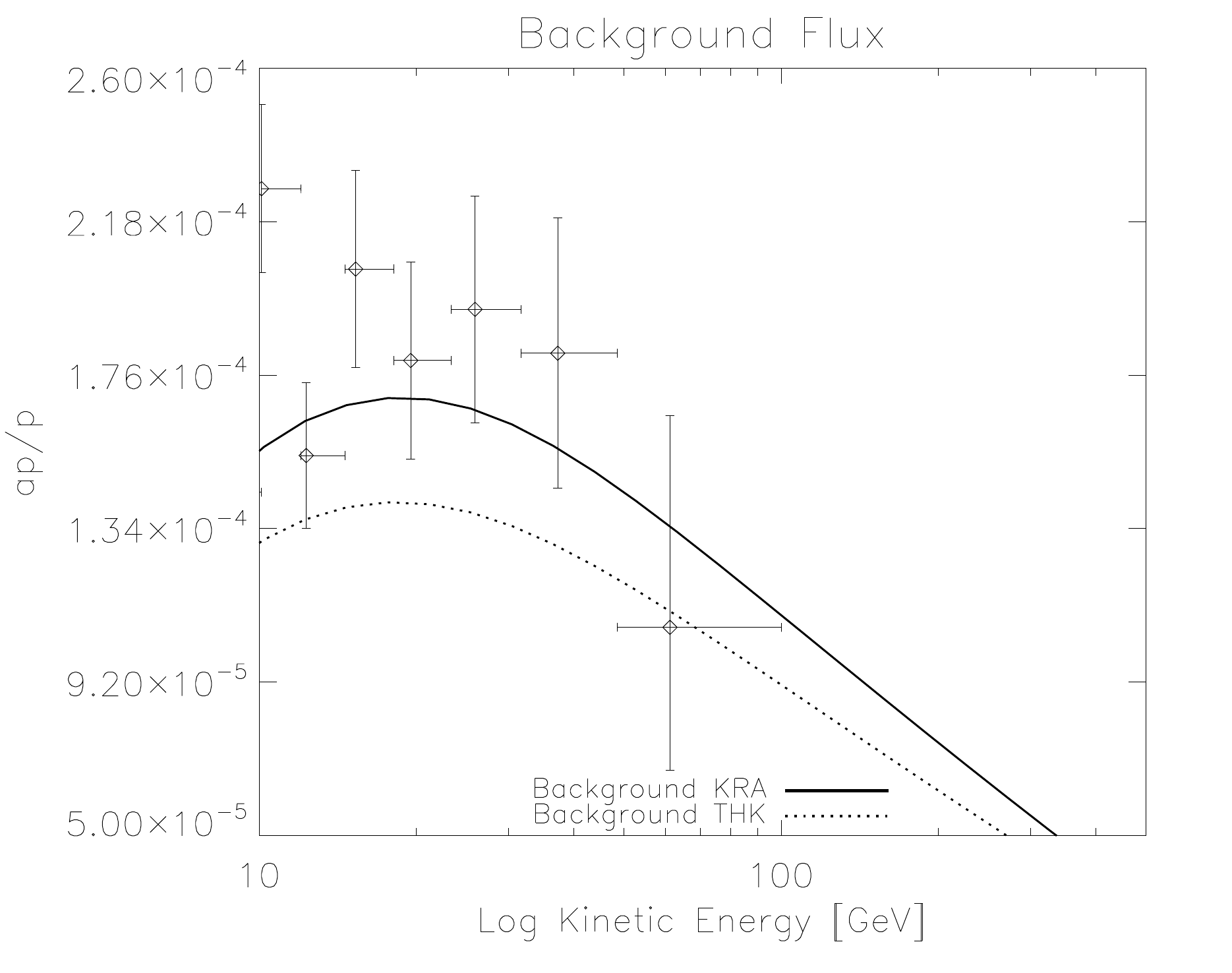}
\caption{Same as Fig.\ref{fig:mock_kra} but with THK propagation model. In the lower panel we show PAMELA data \cite{PamelaAntiprotons} as compared to  the same background curve as in Fig.\ref{fig:mock_kra} for KRA and to the upper panel of this Figure for THK. We keep the same range as in the other panel to facilitate the comparison.
}
\label{fig:mock_thk}
\end{center}
\end{figure*}

To generate the AMS-02 mock data, we have first set the width of the
energy bins based on the detector energy resolution to be \cite{ting}
\be 
\label{resolution} \Delta E / E  = \left( 0.042 (E / \mathrm{GeV}) 
+ 10 \right)\%.
\ee 
The mock data  have as central value of $\bar{p}/p$ the one of the benchmark model in the centre of each bin. 
Uncertainties around each point have been  calculated by summing up in quadrature systematic
and statistical errors for the $\bar{p}/p$ ratio.
The statistical error is approximately given by \cite{Pato:2010ih, Cirelli:2013hv}

\be
\frac{\Delta (\bar{p}/p)^{\rm stat}}{\bar{p}/p} \sim \frac{\Delta N_{\bar{p}}^{\rm stat}}{N_{\bar{p}}} = 
\frac{1}{\sqrt{N_{\bar{p}}}}.
\ee
We have fixed the relative systematic error to be $\Delta N_{\bar{p}}^{\rm syst}/N_{\bar{p}} = 10 \%$. 
Here ${N_{\bar{p}}}$ is the expected number of antiproton events per bin and is  related to the specification parameters of the experiment via the relation
$N_{\bar{p}} = \epsilon \,\, a_{\bar{p}} \,\, \Phi_i \Delta E \Delta t_i$. In particular, we have set the efficiency $\epsilon_i = 1$, the geometrical acceptance of the instrument  $a_{\bar{p}} = 0.2\, \text{$m^2$ sr} $ and a reference operation time $\Delta t_i =  1 \, \, \text{yr}$. The flux $\Phi_i$ is
the $\bar{p}$ flux
in the centre of the bin $i$, while $\Delta E$ is the energy resolution for our binning, as found in Eq. (\ref{resolution}).
Mock data are plotted in Fig.~\ref{fig:mock_kra} for KRA and Fig.~\ref{fig:mock_thk} for THK propagation models.
They extend up to $E_k\simeq400\GeV$; having a higher energy reach would probably improve the discrimination between DM and SNR models.

\begin{figure}
\centering
\subfloat[]{\includegraphics[width=3.1in]{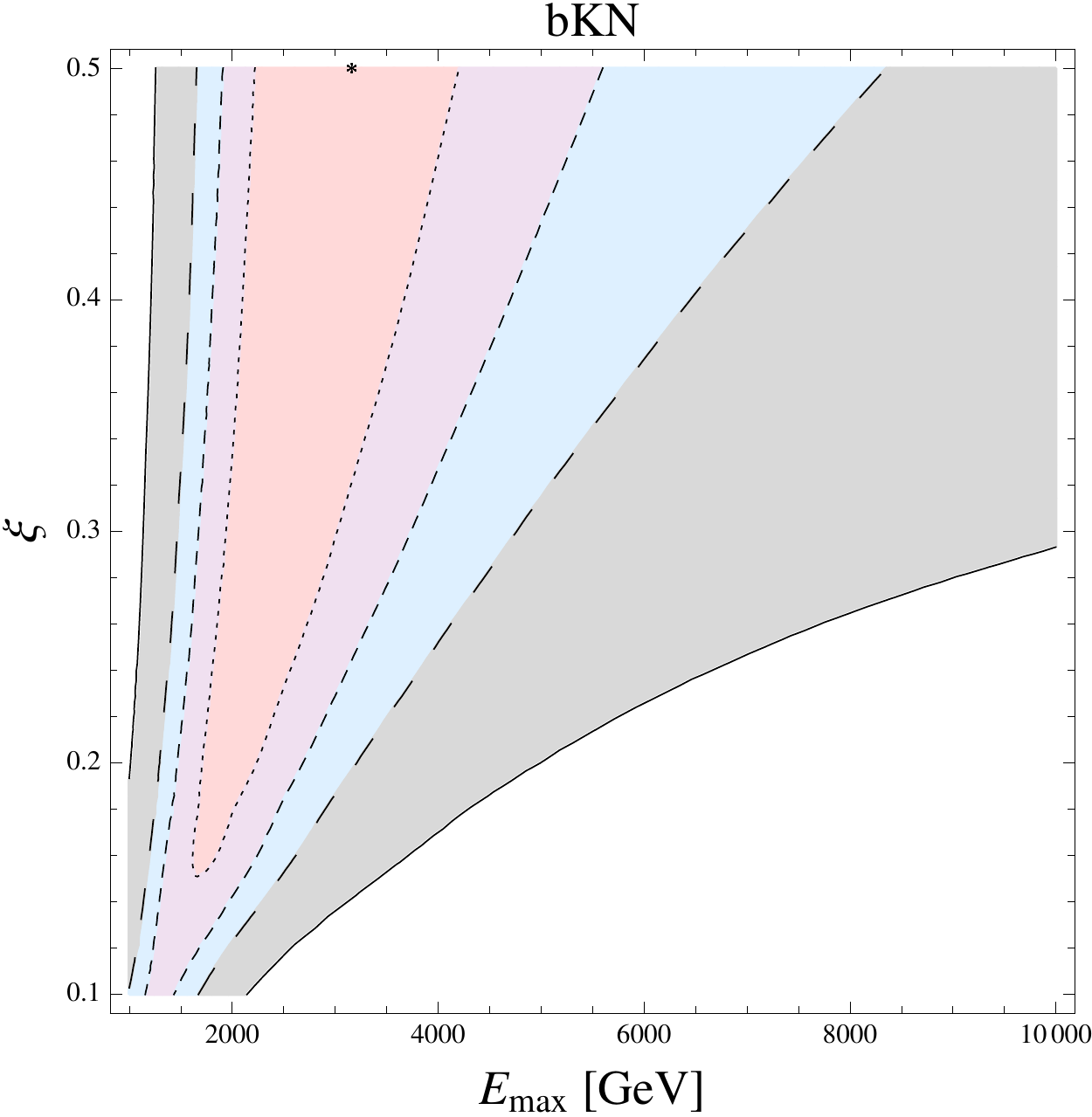}}
\subfloat[]{\includegraphics[width=3.1in]{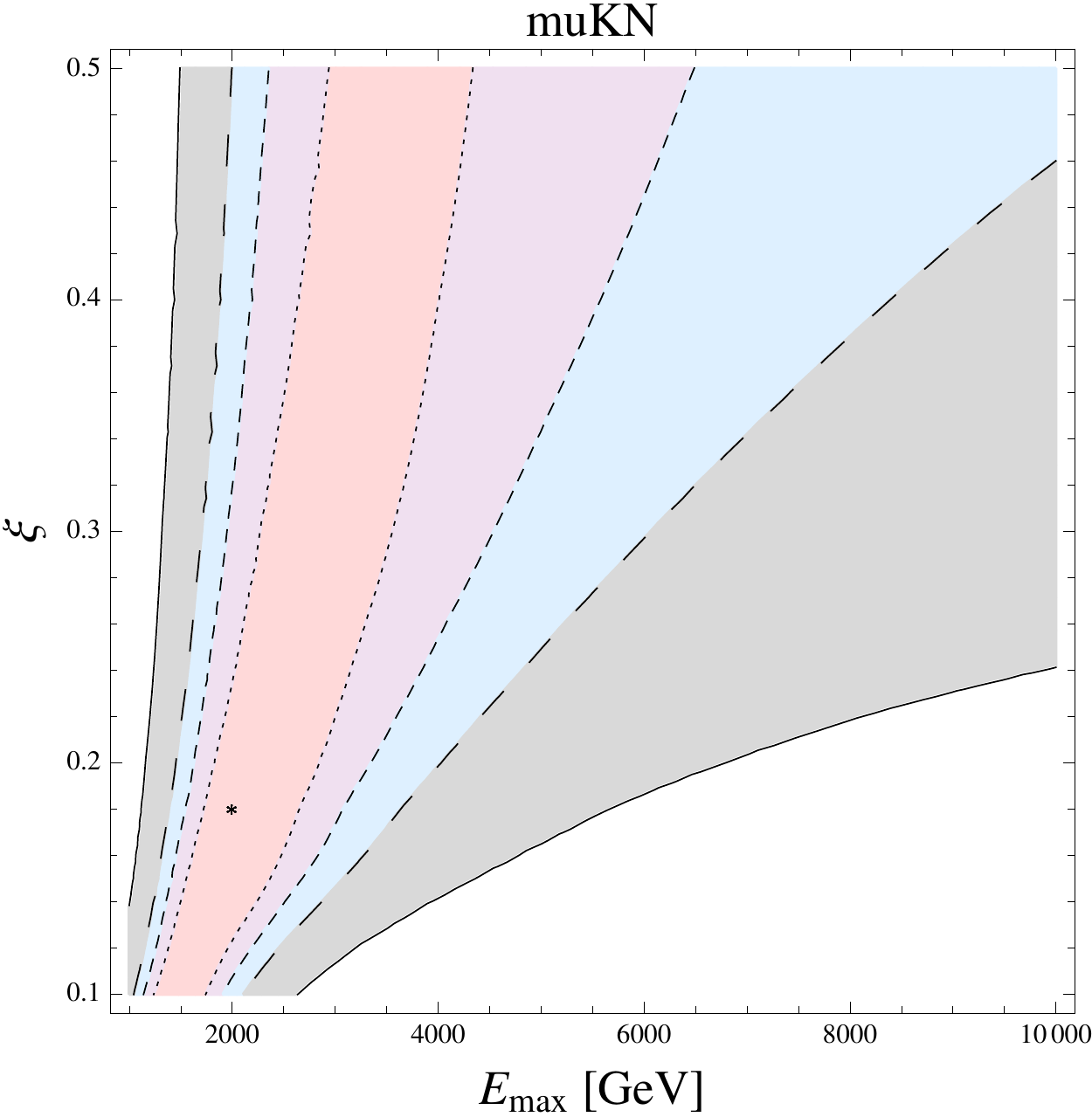}}
\\
\subfloat[]{\includegraphics[width=3.1in]{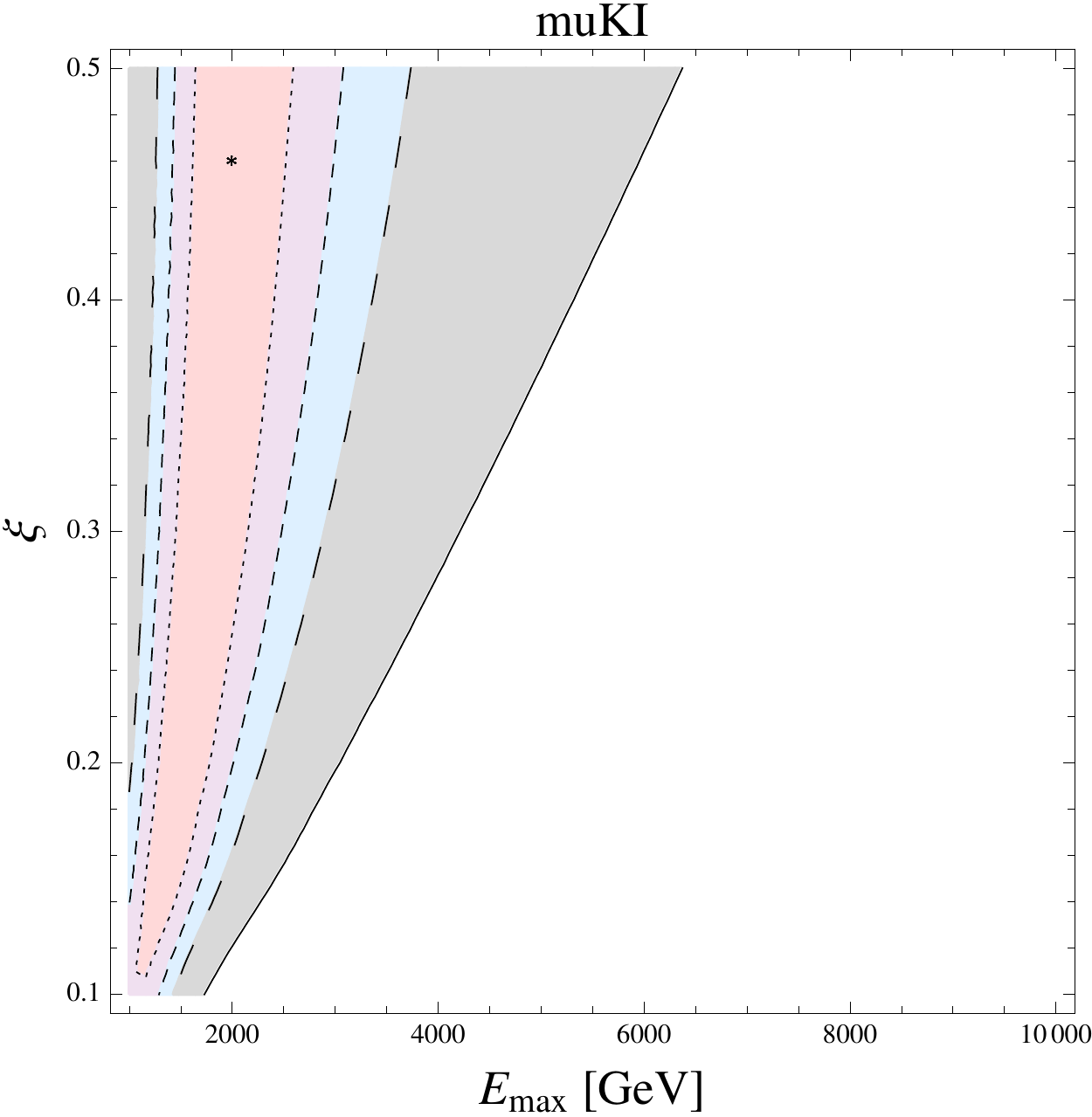}}
\subfloat[]{\includegraphics[width=3.1in]{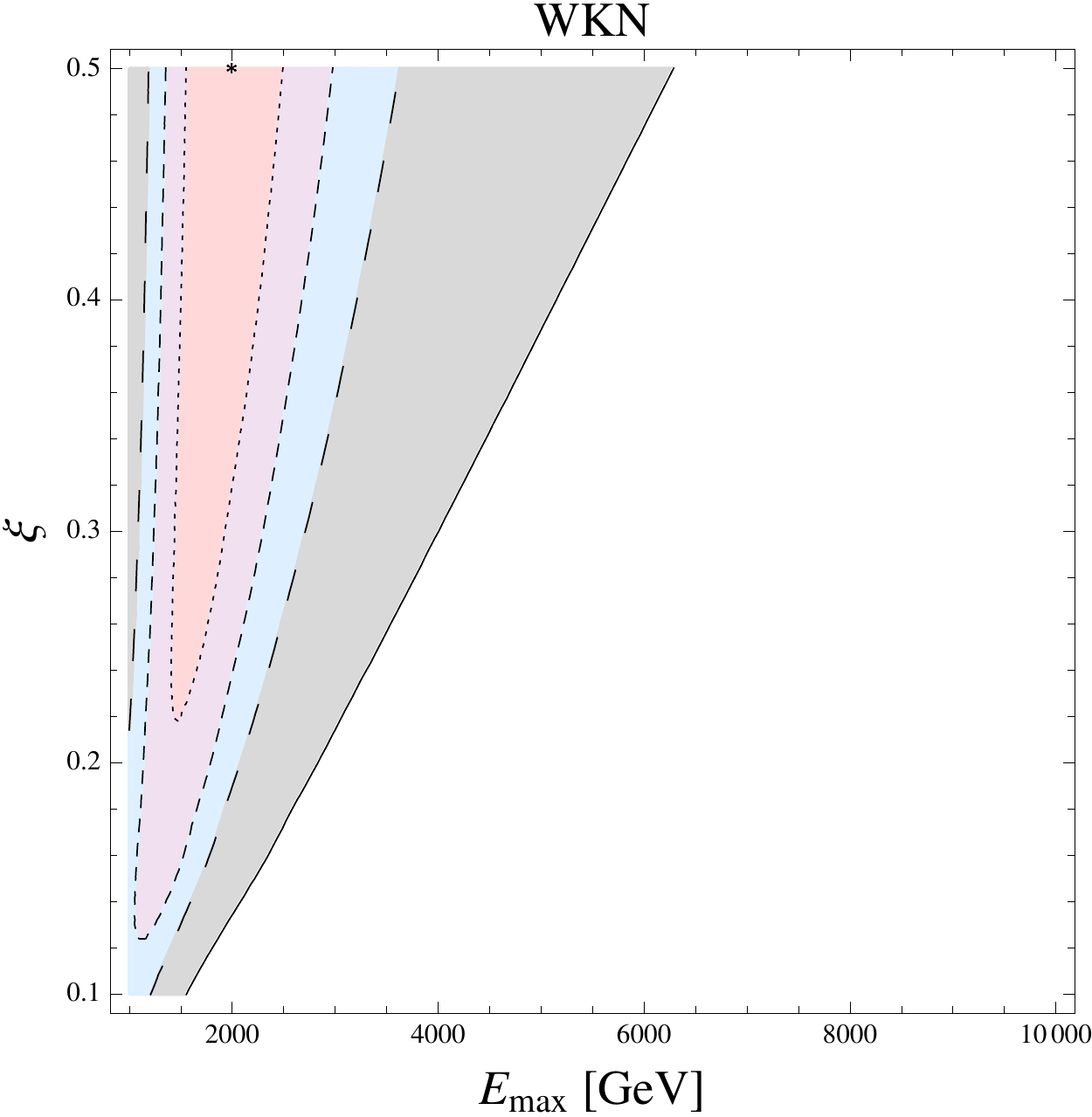}}
\caption{Confidence contours for different DM models with propagation KRA. The names of the models refer to the ones given in Table \ref{tab:result}. Colours indicate 1, 2, 3, 5 $\sigma$ contours. The black dot corresponds to the minimum $\chi^2$ value (relative minimum within the chosen grid). }
\label{fig:contours_kra}
\end{figure}

\begin{figure}
\centering
\subfloat[]{\includegraphics[width=3.1in]{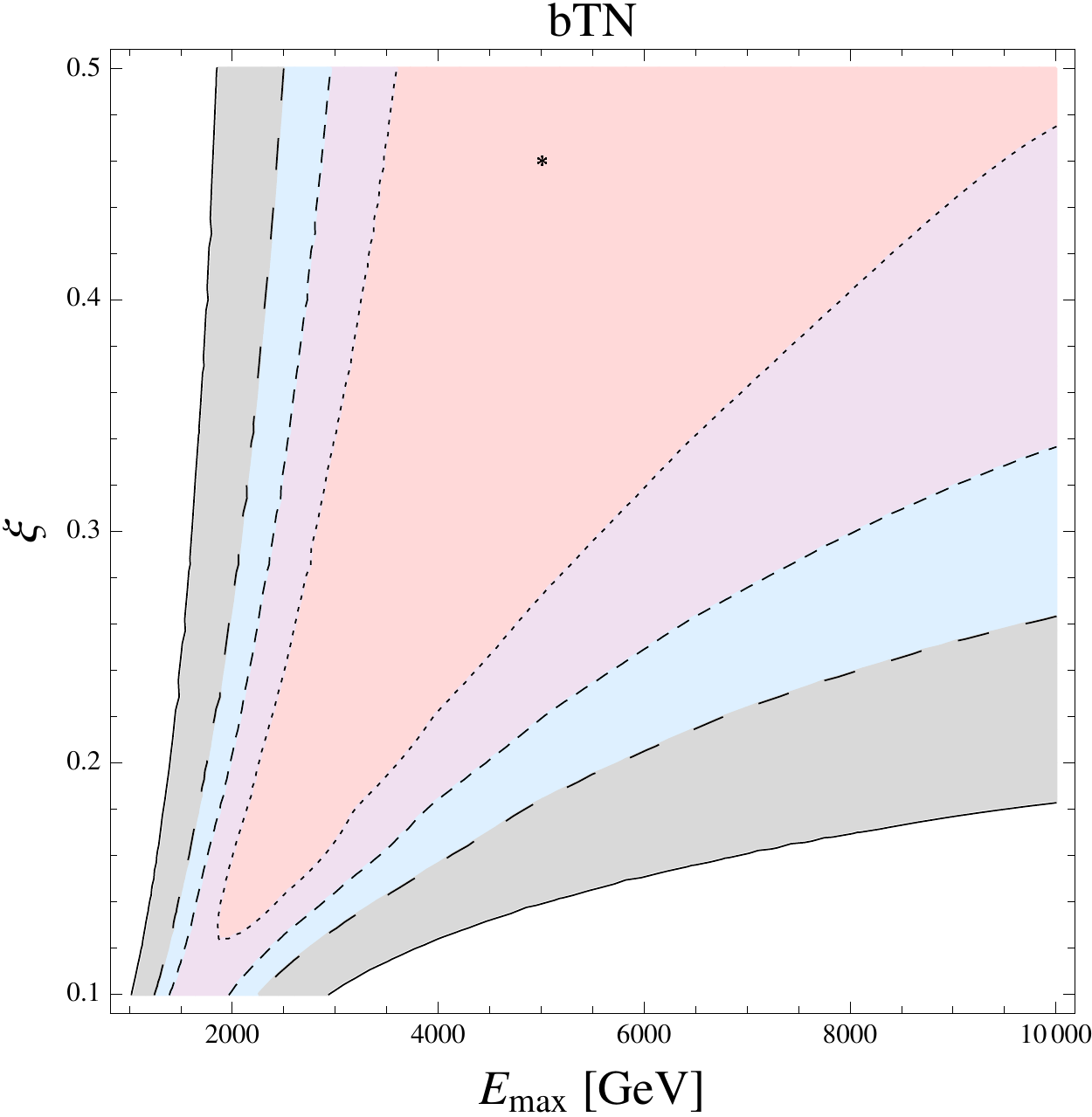}}
\subfloat[]{\includegraphics[width=3.1in]{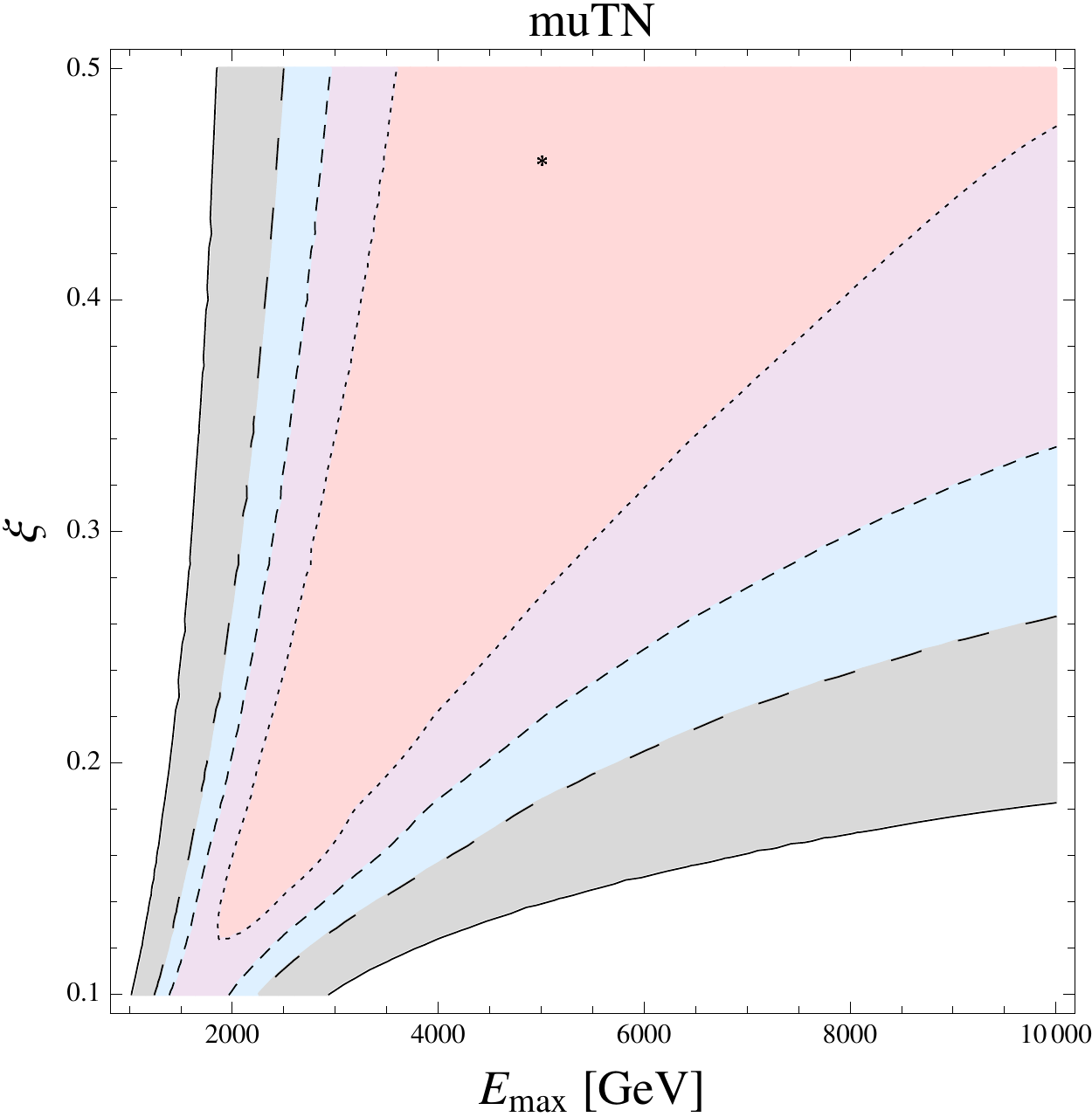}}
\\
\subfloat[]{\includegraphics[width=3.1in]{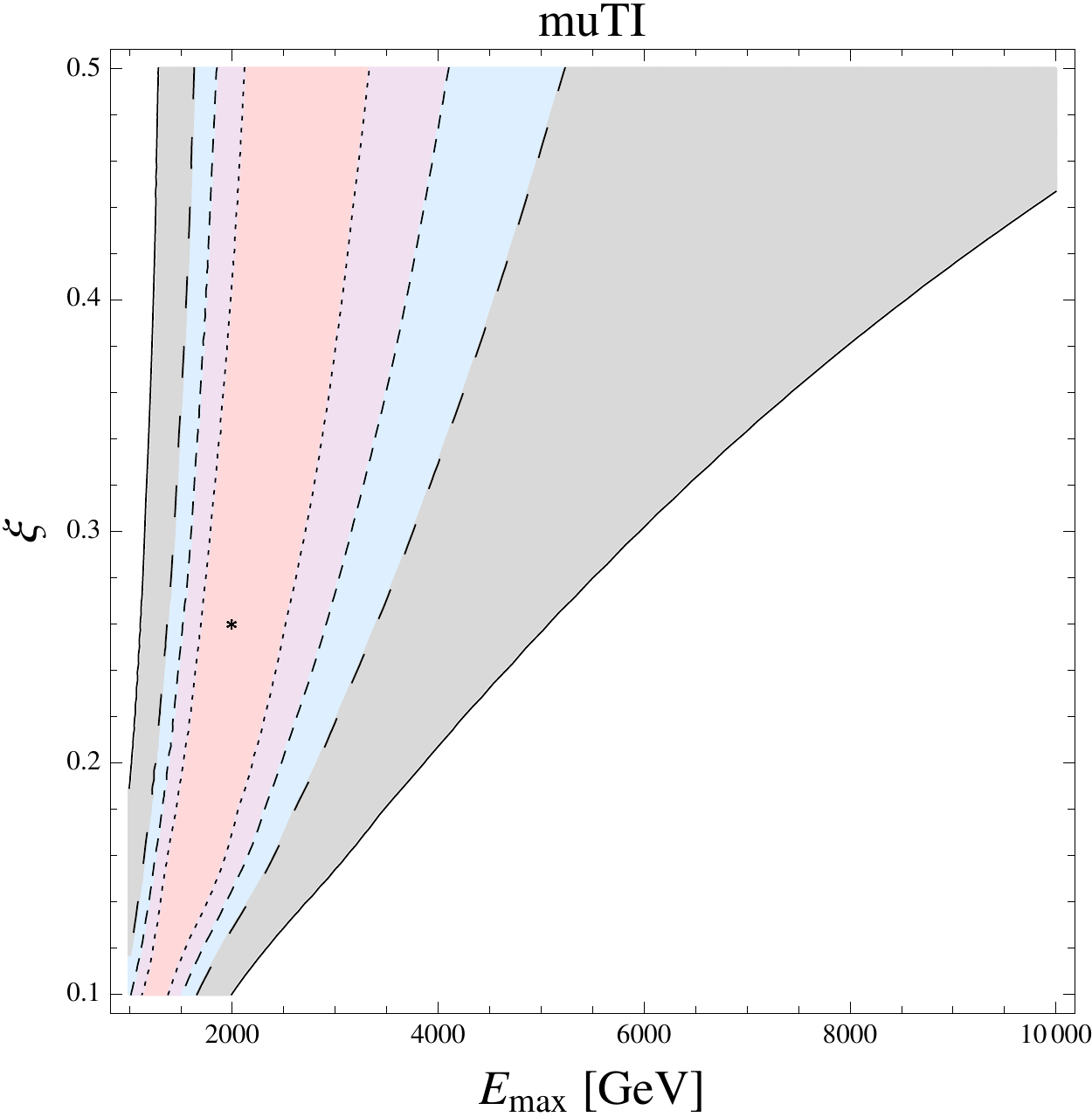}}
\subfloat[]{\includegraphics[width=3.1in]{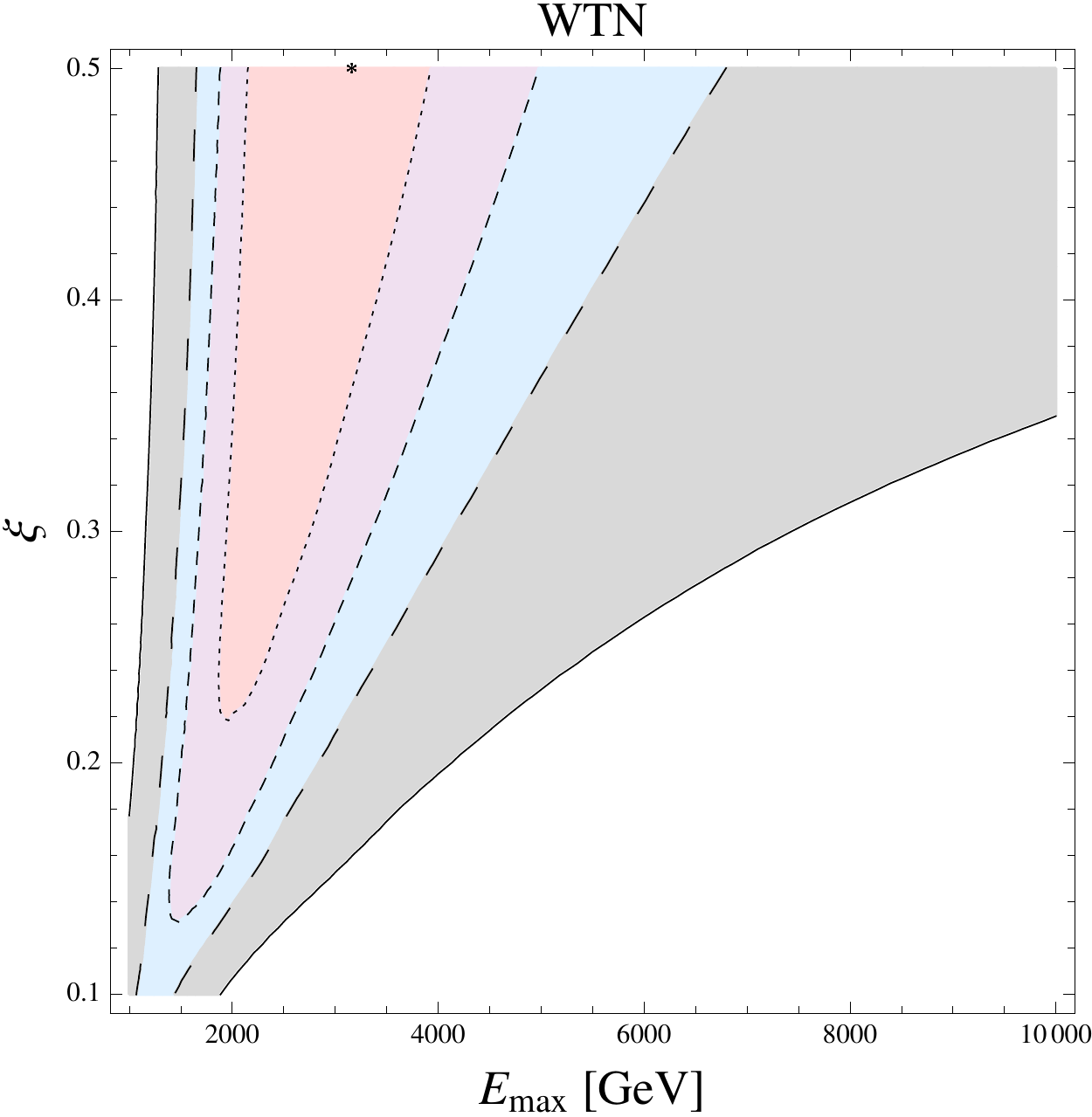}}
\caption{Confidence contours for different DM models with propagation THK. The names of the models refer to the ones given in Tab. \ref{tab:result}. Colours indicate 1, 2, 3, 5 $\sigma$ contours. The black dot corresponds to the minimum $\chi^2$ value (relative minimum within the chosen grid).
}
\label{fig:contours_thk}
\end{figure}

We are now able to quantify  the capability of the
SNR to reproduce possible antiproton fluxes generated  by the DM models (as forecasted for the AMS-02). 
The  SNR fluxes are calculated on the grid of values $(E_{\rm max}, \xi)$.
Confidence contours in the plane $(E_{\rm max}, \xi)$ are shown in Fig. \ref{fig:contours_kra} and Fig. \ref{fig:contours_thk} for all benchmarks DM models in Tab. \ref{tab:result}. Different colours represent $1\sigma$ to $5\sigma$ contours. We have assumed for simplicity a Gaussian distribution. 
Fig. \ref{fig:contours_kra} shows results for the four DM models in Tab. \ref{tab:result} whose propagation follows the KRA prescription. We see that for all annihilation channels ($b, \mu, W$) there can be degeneracy between the corresponding DM model and SNR flux. A point in the grey region indicates that for those choice of $\xi, E_{\rm max}$  the SNR flux is compatible (and therefore degenerate) with mock data based on a DM hypothesis at  $5\sigma$.
In particular, lower values of $E_{\rm max}$ allow for a larger degeneracy in all cases investigated here. The $b$- and $W$-channels seem to prefer larger values of $\xi$ (with relative minimum at the edge of the grid) while the $\mu$-channel has a minimum $\chi^2$ for lower values of $\xi$. Notice though that the tendency towards lower values of $\xi$  disappears when we change DM profile (Fig.  \ref{fig:contours_kra}, panel (c)) or when we change the propagation model, as in (Fig. \ref{fig:contours_thk}, panel (b)). The values of the minimal $\chi^2$ and number of degrees of freedom for all cases is shown in Tab.(\ref{tab:resultchi}) for all models considered in the analysis.

There is indication that some portion of parameter space might be excluded by data on boron to carbon ration, as shown in \cite{cholis_hooper_2014}.
However, we cannot make a direct comparison with the results of this paper because of a different choice of parameters. In particular our case corresponds indeed to $n_{gas} = 2 \cm^{-3}$, $B = 1 \mu G$ and $v = 0.5 \times 10^{-8} \cm/\s$, which can be compared with Fig.3 of their analysis (upper panel) for $K_B = 20$. We are however fixing r = 3.22 as explained in our Section (\ref{sec:SNR}) for consistency with the background spectrum. The paper \cite{cholis_hooper_2014} uses instead r = 4.

\begin{table}[htb]
\begin{center}
\begin{tabular}{|c|c| c| c | c |c | c }
   \hline
Name &      Minimum $\chi^2$      \\
   \hline
   \hline
bKN &    $6.1$      \\
   \hline
muKN &     $6.3$    \\
   \hline
muKI &    $6.7$    \\
   \hline
 WKN &   $21.0$      \\
   \hline
      \hline
 bTN &    $5.6$       \\
   \hline
  muTN  & $5.6$      \\
   \hline
  muTI &  $8.6$      \\
   \hline
  WTN & $8.6$    \\
   \hline
\end{tabular}   
   \caption {$\chi^2$ values for the models considered in this analysis. In all cases the number of degrees of freedom is N = 30 (data points) - 2 (parameters) = 28.
}   \label{tab:resultchi}
\end{center}
\end{table}

\begin{figure}
\centering
\includegraphics[width=3.1in]{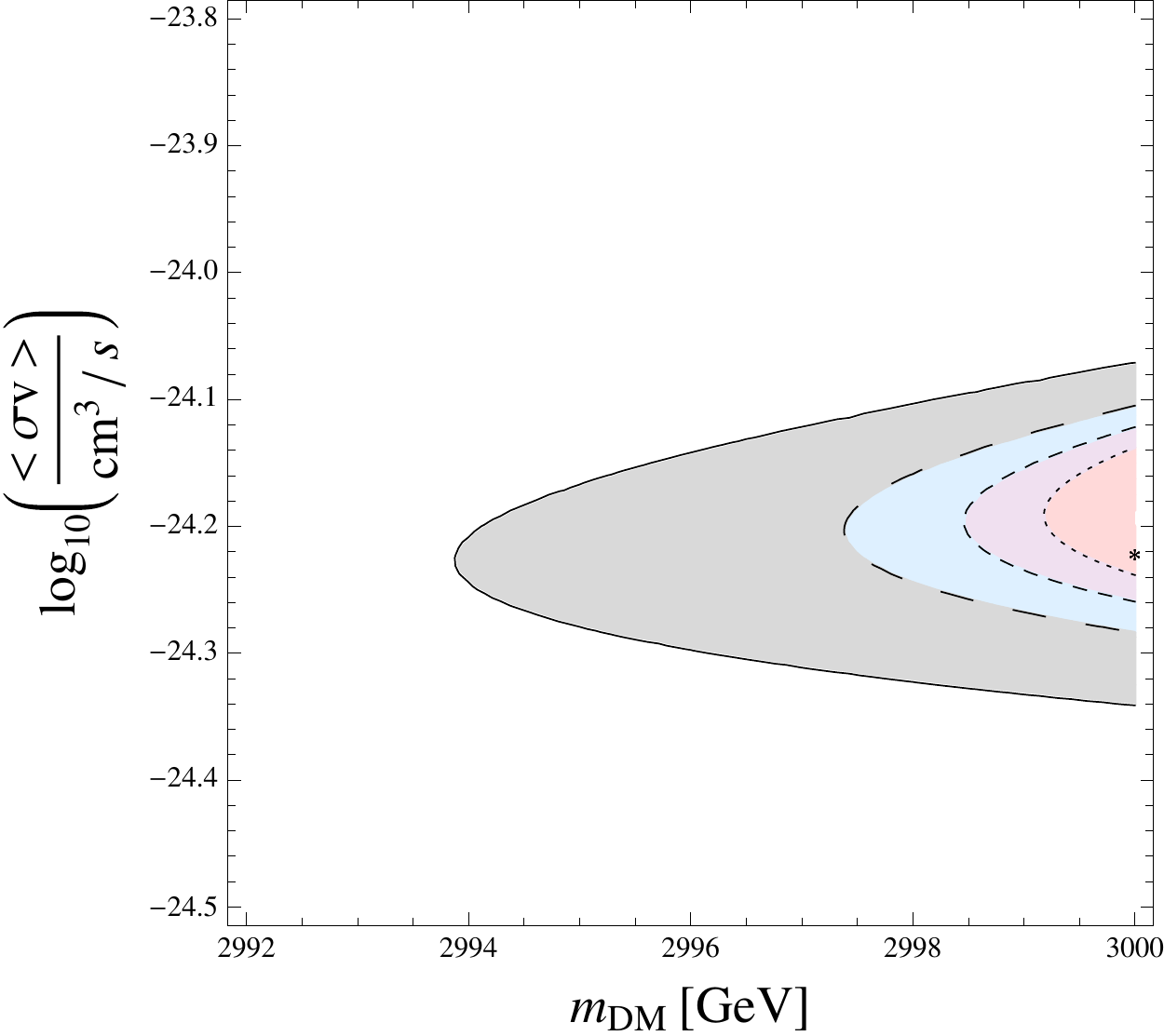}
\caption{Confidence contours in the parameter space ($\langle\sigma v\rangle, M$), for the $ b\bar{b}$ annihilation channel and with KRA propagation model, as obtained fixing $E_{max} = 2500$ GeV and $\xi = 0.14$ in the SNR benchmark model. The (relative) minimum $\chi^2$ within the grid for this case is 9.1 for 28 degrees of freedom. Colours indicate 1, 2, 3, 5 $\sigma$ contours.}
\label{fig_contours_reverse}
\end{figure}

Finally, we have investigated the degeneracy following the inverse logic with respect to the analysis done so far; instead of assuming a DM benchmark model and test whether we can find a combination of $(\xi, E_{max})$ that fit our mock data, we reversed the procedure: 
we first produced a set of mock AMS-02 data through a benchmark SNR model and asked if these data could be interpreted as originated from DM models (using the same propagation model). As expected, also in this case it is possible to find some degeneracy. In Fig.~\ref{fig_contours_reverse} we show an example of such a degeneracy, which, for the chosen SNR benchmark model and DM annihilation channel, peaks around a very small range in mass. This is in agreement with the value found in model bKN. The extension of the degeneracy does not vary much with the annihilation channels.

\section{Conclusions}
\label{sec:conclusions}

Finding  indirect signatures  of DM is certainly one of the main targets of many current experimental efforts. Nevertheless, even in the optimistic case in which a signal above the expected background is found, the most pressing question is whether such a signal can be ascribed to DM annihilation (or decay) beyond any reasonable doubt. 
This is a legitimate question as there are astrophysical sources which can mimic a signal, the best example being pulsars which can generate a  positron excess. In this paper we have
investigated this degeneracy problem focussing our attention on the antiproton signal, in view
of the forthcoming release of data from the AMS-02 collaboration. Indeed, antiprotons may be generated as secondaries accelerated in supernova remnants and we have shown that
a potential signal from DM annihilation can be mimicked by such an astrophysical source.

\section*{Acknowledgments}
We thank Pasquale Serpico for many useful discussions concerning SNR; Carmelo Evoli, Daniele Gaggero and Luca Maccione for many insights about the DRAGON numerical code; Valerio Marra for numerical tips.
ADS acknowledges partial support from the  European Union FP7  ITN INVISIBLES 
(Marie Curie Actions, PITN-GA-2011-289442).
VP acknowledges the  Marie Curie Intra European Fellowship
``DEMO"  within the 7th Framework  Programme of the European Commission and the Transregio TRR33 grant on `The Dark Universe'.

\end{document}